\documentclass[12pt,preprint2]{emulateapj}
\usepackage{pslatex}
\usepackage[T1]{fontenc}
\usepackage[latin1]{inputenc}
\usepackage{subfigure}
\usepackage{amsmath}
\usepackage{graphicx,graphics}
\usepackage{amssymb}
\usepackage[english]{babel}

\newcommand{\be}{\begin{equation}}
\newcommand{\ee}{\end{equation}}
\newcommand{\beqn}{\begin{eqnarray}}
\newcommand{\eeqn}{\end{eqnarray}}
\newcommand{\lap}{\lesssim}
\newcommand{\gap}{\gtrsim}

\newcommand{\msun}{M_\odot}

\newcommand{\beq}{\begin{equation}}
\newcommand{\eeq}{\end{equation}}
\newcommand{\mh}{M}

\def\gap{\;\rlap{\lower 2.5pt
 \hbox{$\sim$}}\raise 1.5pt\hbox{$>$}\;}
\def\lap{\;\rlap{\lower 2.5pt
   \hbox{$\sim$}}\raise 1.5pt\hbox{$<$}\;}

\shorttitle{Loss Cone Refilling}
\shortauthors{Merritt and Wang}
\begin{document}

\title{Loss Cone Refilling Rates in Galactic Nuclei}

\author{David Merritt}
\affil{Department of Physics, Rochester Institute of Technology, 
Rochester, NY 14623}

\author{Jianxiang Wang}
\affil{Department of Physics and Astronomy, Rutgers University, New Brunswick, 
NJ 08903}

\begin{abstract}
A gap in phase space is opened up by a binary supermassive
black hole as it ejects stars in a galactic nucleus.
This gap must be refilled before the single black hole
that subsequently forms can disrupt or accrete stars. 
We compute loss cone refilling rates for a sample of
elliptical galaxies as a function of the mass ratio of the
binary that preceded the current, single black hole.
Refilling times are of order $10^{10}$ yr or longer in
bright elliptical galaxies.
Tidal flaring rates in these galaxies might be much lower
than predicted using steady-state models.

\end{abstract}


\section{Introduction}

The loss cone of a black hole (BH) at the center of a galaxy is
defined as the set of orbits that intersect the BH,
or that pass within some distance of its center.
For instance, the tidal disruption loss cone consists 
of orbits with pericenters below $r_t$,
the radius at which tidal forces from the BH  would
disrupt a star.
Stars on such orbits are removed in a single
orbital period or less, and subsequent feeding of 
stars to the BH requires
a re-population of the loss cone, which is typically
assumed to be driven by gravitational encounters between
stars.

Classical loss-cone theory
\citep{bw-76,ls-77,ck-78}
was directed toward understanding the observable 
consequences of massive BHs at the centers of globular clusters.
Globular clusters are many relaxation times old,
and this assumption was built into the theory,
by requiring the stellar phase space density near the BH to have 
reached an approximate steady state
under the influence of gravitational encounters.
In a collisionally-relaxed cluster, the density of stars
near the BH has the Bahcall-Wolf (1976) ``zero-flux''
form, $\rho\propto r^{-7/4}$, and the dependence of
$f$ on $J$, the stellar angular momentum per unit mass, 
near the loss-cone 
boundary is described by the Cohn-Kulsrud (1978) boundary
conditions.
The feeding rate is determined by the gradients of
$f$ with respect to $J$ at the loss cone boundary and 
by the normalization of $f$ at values of $J$ far from the 
loss cone, i.e. by the stellar density.

Galactic nuclei differ from globular clusters in that the
relaxation time is often very long, typically in excess
of $10^{10}$ yr \citep{faber-97}.
One consequence is that the stellar density profile
near the BH need not have the Bahcall-Wolf $r^{-7/4}$
form.
But the fact that galactic nuclei are not collisionally relaxed
also has implications for the more detailed form of the
phase space density near the loss cone boundary,
and hence for the feeding rate.
In one widely discussed model for the formation
of galactic nuclei \citep{bbr-80}, a binary BH forms following the
merger of two galaxies.
The binary ejects stars on orbits
such that $J<J_{bin}\approx \sqrt{2GM_{12}a_h}$,
with $M_{12}$ the mass of the binary and $a_h\approx G\mu/4\sigma^2$
the binary's semi-major axis at the time of its
formation; $\mu=M_1M_2/M_{12}$ is the reduced mass and
$\sigma$ is the stellar velocity dispersion.
Since $a_h\gg r_t$, there will be a 
gap in angular momentum space around the single
BH, of mass $M_\bullet=M_{12}$,  that subsequently forms, 
corresponding to stars with $J\lap J_{bin}$ 
that were ejected by the binary.
Before the single BH can begin to 
consume stars at the steady-state rate, 
this gap needs to be re-filled.
This argument suggests that the feeding rates of
BHs in galaxies that formed via mergers might be
much lower than predicted by application of the
steady-state theory.
Here we solve the time-dependent equations describing
the evolution of the stellar phase-space density
around a BH in a galactic nucleus and compute
the time required for the steady-state feeding
rate to be reached.

\section{Approximate Timescales}

Diffusion of stars into a single BH is dominated by
scattering onto low-angular-momentum orbits \citep{fr-76}.
The scattering time is
\beq
T_\theta(r) \approx \theta(r)^2T_r(r)
\eeq
where $\theta(r)$ is the angle within which a star's velocity
vector must lie if it is in the loss cone,
and $T_r$ is the relaxation time,
\beq
T_r(r) = {\sqrt{2}\sigma(r)^3\over\pi G^2m_\star\rho(r)\log\Lambda},
\eeq
with $m_\star$
the stellar mass and $\log\Lambda\approx 15$ the Coulomb logarithm.
The square-root dependence of $\theta$ on $T_\theta$
reflects the fact that entry into the loss cone is a diffusive process.
If the loss cone was initially emptied of all stars with
pericenters below some radius $r_0$, 
the time to diffusively refill this angular momentum gap is
\beq
T_{gap} \approx {r_0\over r}T_r.
\label{eq:tgap}
\eeq
Equation \ref{eq:tgap} assumes that 
$r_t \ll r_0 \lap r_h$ with $r_h=GM_{12}/\sigma^2$ the
influence radius of the single (coalesced) BH \citep{fr-76}.
Until the angular momentum gap is filled, the rate of supply of stars
to the BH's capture sphere will be much less
than its steady-state value.

In the binary coalescence model,
most stars with pericenters
\beq
r_p \lap Ka_h=K{G \mu\over 4\sigma^2}
\label{eq:rp}
\eeq
will have been ejected prior to coalescence
by the gravitational slingshot mechanism \citep{saslaw-74},
with $K$ a constant of order unity.
This ejection will occur even if the binary's main
source of angular momentum loss is torques from gas clouds,
as long as $|a/\dot a|$
exceeds stellar orbital periods near the BHs.
Setting $r_0\approx r_p$ and $r\approx r_h$ (since
most of the scattering into the single BH takes place from
radii near $r_h$), we find
\beq
T_{gap}\approx {K\over 4}{q\over (1+q)^2} T_r(r_h)
\label{eq:tgap2}
\eeq
with $q=M_2/M_1\le 1$.
For $q=1$, $T_{gap}\approx T_r/16$, and for 
$q\ll 1$, $T_{gap}\approx (q/4)T_r$.
These times are short compared with $T_r$ but
still in excess of $10^{10}$ yr for the luminous
galaxies that are most likely to have experienced
mergers.

The principal uncertainties in this estimate of the 
refilling time are the value of $K$, and the fact that stars
are scattered into the loss cone from a range of distances
with different values of $T_r$.
We address these issues by more careful calculations in the
following sections.

\section{Creation of a Phase-Space Gap}

The gap is created by the binary BH 
when its separation reaches $\sim a_h$.
The definition given above for $a_h$ is difficult to apply
to galaxies in which $\sigma$ is a function of radius.
We therefore adopt an alternative definition in terms of
$r_h$:
\begin{equation}
a_h = {\mu\over M_{12}}{r_h\over 4} = {q\over (1+q)^2}{r_h\over 4},
\end{equation}
with $r_h$ defined as the radius in the unperturbed 
galaxy containing a mass in stars equal to twice $M_{12}$.
These definitions are equivalent to the standard
ones $r_h=GM_{12}/\sigma^2$, $a_h=G\mu/4\sigma^2$
in a singular-isothermal-sphere nucleus but can also 
be applied to nuclei in which $\sigma$ is a function of radius.

Once it forms, the binary shrinks as it ejects stars
that pass within a distance$\sim a(t)$ of the binary's
center of mass.
The coupled evolution of the stellar fluid and the binary
is complicated due to processes like re-ejection, the
repeated interaction of (non-escaping) stars with the binary,
and by the fact that the binary's cross-section is changing
with time \citep{merritt-04}.

To model this process, and to compute the size of the gap
in terms of $a_h$ as we have defined it,
we carried out a set of Monte-Carlo simulations.
Scattering experiments were first carried out for an
isolated, circular-orbit binary as a function of the parameters
$(R_p,V_p)$, the distance and velocity at closest
approach to the binary that a test particle would 
have if the binary were replaced by a point of mass
$M_{12}$ at the binary's center of mass.
The changes in energy and angular momentum,
$\Delta E$ and $\Delta J$, experienced during
the encounter were recorded and the calculation repeated for
a large number ($\sim 10^6$) of different $(R_p,V_p$) values.

In the second step, a set of Monte-Carlo positions and
velocities were generated for a set of stars drawn from
an isotropic distribution function describing the stars
in the unperturbed galaxy.
The gravitational potential was fixed as
$\Phi(r) = \Phi_\star(r) - GM_{12}/r$ 
with $\Phi_\star(r)$ the contribution to the potential
from the stars.
The stellar orbits were integrated forward in time,
and stars were given kicks in energy and angular momentum
whenever they passed within a distance $r_{crit}(t)=3a(t)$
of the galaxy center.
The kicks were drawn randomly from the stored values of
$\Delta E$ and $\Delta J$; the star's orbit in the 
full potential was related to the Keplerian orbit in the
scattering experiments by equating the pericenter
distance $r_p$ and velocity at pericenter $v_p$
of the orbit about the galaxy's center with $R_p$ and $V_p$.
Immediately after a kick, a star was placed on an
outgoing orbit on the sphere $r=r_{crit}$ with a velocity
computed using its new values of $E$ and $J$.
At the same time, the binary's energy was reduced
by $m_\star\Delta E$ and its semi-major axis
correspondingly changed.
Possible changes in the eccentricity of the binary's orbit
were ignored.
Stars with initial apocenters within $a(0)$ were flagged 
and removed;
in a real galaxy, such stars would have been ejected during
the binary's formation.
The stellar potential was taken to be that of a
\cite{dehnen-93} model; 
initial positions and velocities
of the stars were generated from the unique isotropic
distribution function $f(E)$ that reproduces the Dehnen
mass distribution in the combined potential of the stars
and the central point mass \citep{tremaine-94}.
The parameters of the Monte-Carlo integrations were
$q$, $M_{12}/M_{gal}$, $a(0)$, and
$\gamma$, the central logarithmic density slope of the
Dehnen model.

\begin{figure}
\plotone{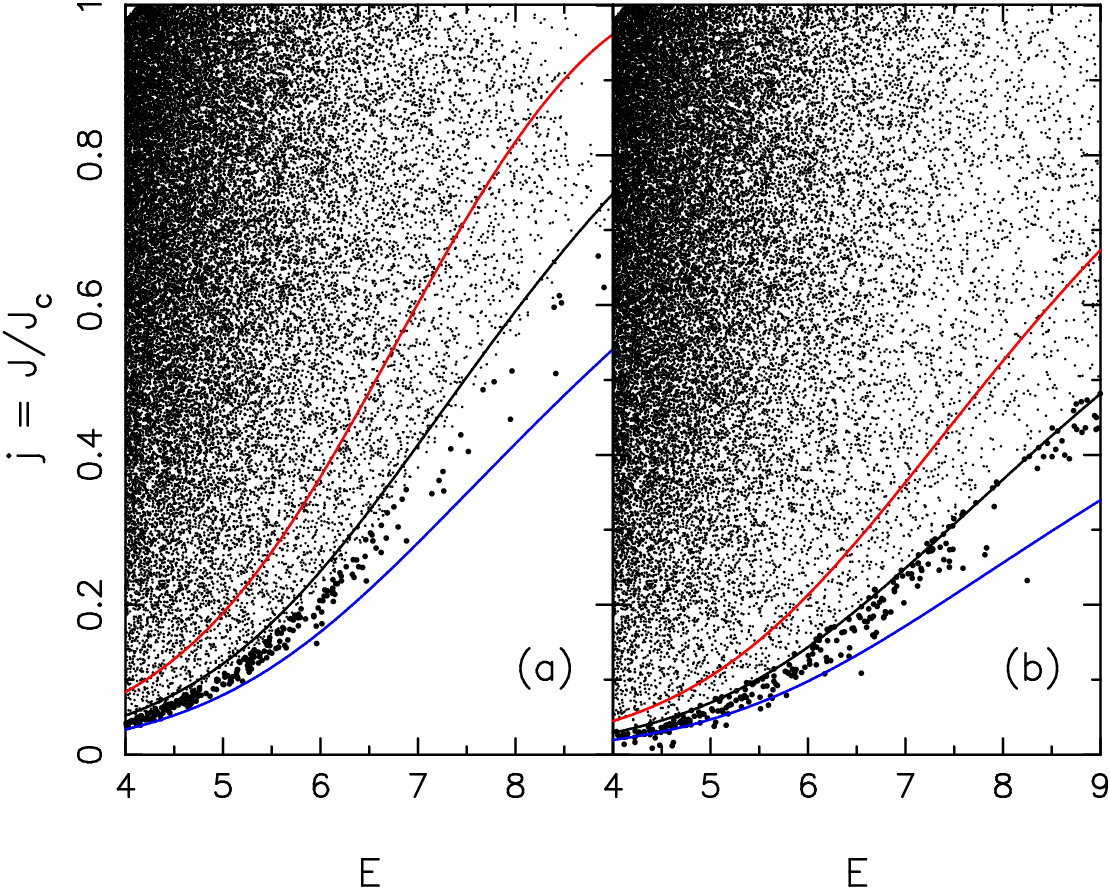}
\caption{
Lindblad diagram of stars at the final time step
in the Monte-Carlo experiments, with $q=1$ (a) and
$q=1/8$ (b).
Additional parameters are given in the text.
Larger dots are stars that are still on orbits
that intersect the binary.
Lines show $j_{gap}$ for $K=0.5$ (blue), $1$ (black) and $2$ (red).
\label{fig:snap}}
\end{figure}

Figure~\ref{fig:snap} illustrates the gap created
in two integrations with $\gamma=2$ and $M_{12}/M_{gal}=0.001$,
and with $q=(1,1/8)$.
The plots show the phase-space distribution at a time
of $\sim 4$ in units such that $G=M_{gal}=r_D=1$
with $r_D$ the Dehnen-model scale length.
The larger circles are stars that are still interacting
with the binary, i.e. stars with $r_p<r_{crit}$.
Most of these stars will eventually be ejected but some
will remain in the galaxy as the binary ``shrinks away''
beneath them.
The three curves in Figure~\ref{fig:snap} are
\begin{equation}
J^2 = 2Ka_h\left[E-\Phi(Ka_h)\right] \equiv J_{gap}^2
\label{eq:jgap}
\end{equation}
with $K=(0.5,1,2)$; $J_{gap}$ is the angular momentum of
a star with pericenter $K a_h$.
There is a sharp drop in phase space density at
$J\approx J_{gap}$, $K\approx 1$.
This relation was found to hold also for other
values of $\gamma$ and $q$.
The location of the gap was found to be only 
weakly dependent on the choice of $a(0)$.
For the integrations of Figure~\ref{fig:snap},
$a(0)\approx 2 a_h$, and increasing $a(0)$ by a factor
of $10$ caused $J_{gap}$ to increase by only
$\sim 50\%$.
We conclude that equation (\ref{eq:jgap}) with $K=1$
is a good representation of the phase-space gap
produced by a binary BH.

\section{Refilling the Gap}

In the Monte-Carlo experiments described above, the binary decay
stalls after all the stars on orbits intersecting the
binary have been ejected.
We assume that some other mechanism, e.g. torques
from gas in an accretion disk, then acts to
extract angular momentum from the binary,
eventually bringing the two components close enough
together that emission of gravitational radiation
can induce complete coalescence.

\begin{figure}
\plotone{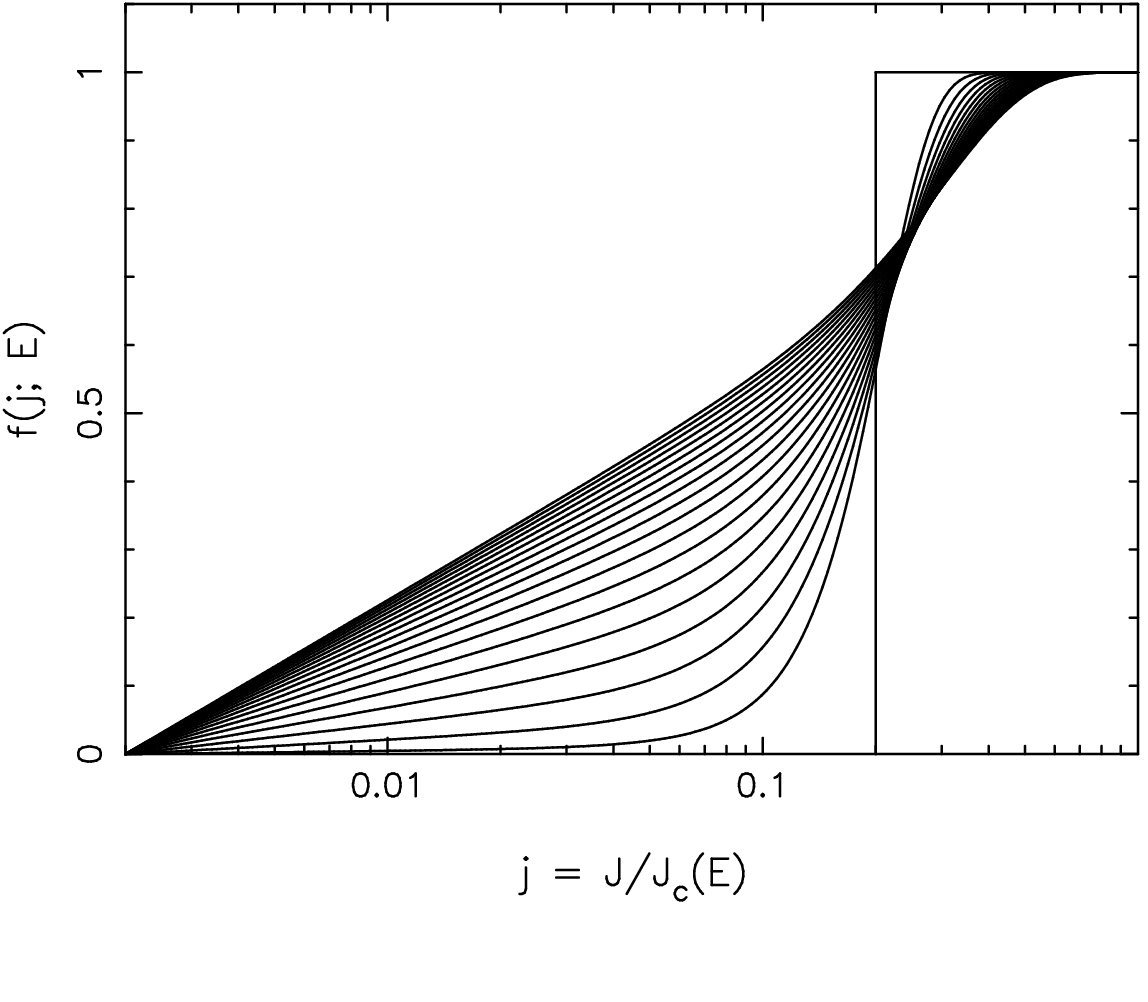}
\caption{
Solution to equation (\ref{eq:fp}) with the initial and boundary
conditions given in the text, at 
dimensionless times 
$\tau=0,1\times 10^{-3},2\times 10^{-3},...,4\times 10^{-2}$
where $\tau=\mu(E)t\approx t/T_r$.
\label{fig:singlee}}
\end{figure}

The phase-space gap opened up by the binary will
then gradually refill on the time scale $T_{gap}$.
We modelled this evolution using the time-dependent
Fokker-Planck equation.
The following simplifying assumptions were made.
(1) Changes in stellar {\it energy} due to encounters
were ignored, i.e., we assumed that $T_{gap}\ll T_r$
(cf. equation \ref{eq:tgap2}).
(2) Relevant stars are those with $R\equiv J^2/J_c(E)^2\ll 1$,
where $J_c(E)$ is the angular momentum of a circular orbit of energy $E$.
Under this assumption, the low-$R$ form of the velocity diffusion coefficient
may be used.
(3) The diffusion in $J$ is slow compared with orbital
periods so that the orbit-averaged form of the Fokker-Planck
equation can be used.
We justify this assumption below.
Given these assumptions, the evolution of the stellar
phase space density $f(J,t;E)$ obeys \citep{mm-03}
\begin{equation}
{\partial f\over\partial\tau} = {1\over 4j}
{\partial\over\partial j}
\left(j{\partial f\over\partial j}\right)
\label{eq:fp}
\end{equation}
where $j\equiv J/J_c(E)=R^{1/2}$ is a dimensionless angular momentum
variable and $\tau\equiv\alpha(E)t$ is a dimensionless time, with
$\alpha(E)$ the orbit-averaged diffusion coefficient at energy
$E$:
\begin{equation}
\alpha(E) = {1\over P(E)} \oint {dr\over v_r} 
\lim_{R\rightarrow 0} 
{\langle\left(\Delta R\right)^2\rangle \over 2R}.
\end{equation}
and $P(E)$ the period of a radial orbit of energy $E$.
Hence $\tau\approx t/T_r(E)$.
Equation (\ref{eq:fp}) can be integrated forward in time
given boundary conditions at $j=0$ and $j=1$
and initial conditions $f(j,t=0;E)$.
Figure~\ref{fig:singlee} shows the result if $f(j,t=0)$ has
the form
\begin{equation}
f=0,\ \  j\le 0.2;\ \ \ \  f=1,\ \  0.2<j<1 .
\end{equation}
These are the initial conditions, at one energy,
created by a binary BH with $j_{gap}=0.2$
in an initially isotropic nucleus.
The boundary conditions are
\begin{equation}
f(j\le j_{lc},t) = 0;\ \ \ \ {\partial f\over\partial j}|_{j=1}=0
\end{equation}
with $j_{lc}$ the scaled angular momentum of a capture orbit;
in Figure~\ref{fig:singlee}, $j_{lc}=0.02$.
The flux of stars into the BH is given by
\begin{equation}
{\cal F}(t;E) = -{d\over dt}\left[\int_{j_{lc}}^1N(j,t;E)dj^2\right] 
= {\alpha\over 2}{\partial N\over\partial\log j}|_{j_{lc}} 
\label{eq:flux}
\end{equation}
with $N(E,R,t)dEdR=4\pi^2P(E)J_c^2(E)f(j^2,t;E)dEdR$
the number of stars in the integral-space element $dEdR$.

The orbit-averaged assumption breaks down sufficiently far from
the binary, in the ``pinhole'' regime \citep{ls-77}.
In the loss cone defined by a {\it single} BH, the breakdown occurs
at binding energies less than $E_{crit}$
where
\begin{equation}
q_{lc}(E)\equiv {\alpha(E)P(E)\over R_{lc}(E)} > 1.
\end{equation}
For $E\lap E_{crit}$,  stars can wander into and out of the loss
cone in a single orbital period and the loss cone remains 
nearly full in spite of capture.
Typically $r_{crit}\approx r_h$, the BH's influence radius
\citep{fr-76}.
In the case of the angular momentum gap
created by a {\it binary} BH, 
the orbit-averaged assumption holds
initially even for $r\gg r_{crit}$, since stars need to diffuse
over the full angular momentum gap before they can be disrupted.
Breakdown of the orbit-averaged approximation only occurs
much farther from the center where
\beq
q_{gap}(E) \equiv {\alpha(E)P(E)\over R_{gap}(E)} > 1.
\eeq
\cite{ck-78} showed that the steady-state form of $f$ near
the loss cone was $f\sim\ln(j/j_0)$ and derived
an expression for $j_0=j_0(E)$; beyond $r_{crit}$,
$j_0\approx 0$, i.e. the loss cone is nearly full.
We adopted the same expression for $j_0$ here and
set $f=0, j\le j_0$ at all times.
While not strictly correct in the case of a time-dependent
loss cone, this boundary condition gives the correct
result as $t\rightarrow\infty$, and the error in the
inferred loss-cone flux at earlier times should be very
small since so little of the galaxy is in the full-loss-cone
immediately after the binary forms.

We applied this prescription to compute the evolution
of $f$ and ${\cal F}$ in the ``core'' galaxies from Paper I.
For each galaxy, the function $\alpha(E)$ was computed
on a grid of energies and stored.
Equation (\ref{eq:fp}) was then integrated forward 
for the appropriate dimensionless time at each $E$ value
and the flux computed from equation (\ref{eq:flux}).
We used the same values for the BH mass $\mh$ and for $r_t$
as in Paper I (only the ``$\mh-\sigma$'' values for $\mh$ 
from that paper were used.)
We tried three different values for the mass ratio of the binary 
that was assumed to have formed the gap:
$q\equiv M_2/M_1=(0.1,0.3,1)$, and $j_{gap}$
was computed from equation (\ref{eq:jgap}) for various
values of $K$; following the discussion above, we focus here
on the results obtained with $K=1$.
At the end of the integration, the galaxy was assumed to
be in its currently-observed state, for which we know
$\overline{f}(E)$, the $R$-averaged distribution function.
The correctness of the integrations was tested by
verifying that the asymptotic flux was equal to the
steady-state values computed in Paper I.

\section{Results}

\begin{figure}
\plotone{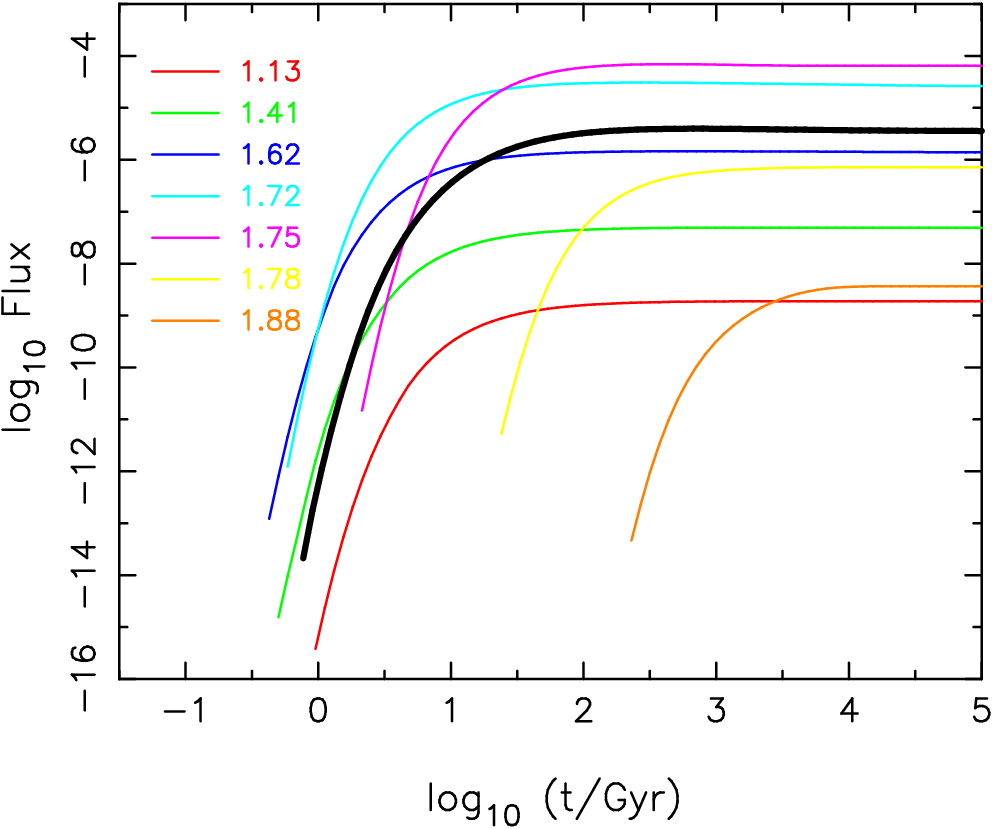}
\caption{
Dependence of the loss-cone flux on energy and time in the
galaxy NGC 4168 assuming a binary mass ratio of $q=0.1$.
Curves are labelled by their binding energy; the value of the gravitational
potential at $r=r_h$ in this galaxy is $1.76$.
Thick black curve is the total flux, in units of stars per year.
\label{fig:fluxet}}
\end{figure}

Figure~\ref{fig:fluxet} shows the energy-dependent evolution of the
loss-cone flux in one of the sample galaxies, NGC 4168,
for $K=1$ and $q=0.1$.
Stars with binding energies near $\Phi(r_h)$ are the first
to be scattered into the BH, followed by stars with energies
slightly above or below $E_h$.
The fact that very little of the flux comes from stars
with energies very different from $E_h$ validates our approximate
treatment of the loss-cone boundary condition in the full-loss-cone
regime.
The total flux reaches $(1\%,10\%,50\%,90\%)$ 
of its steady-state value in  a time of 
$\sim 4.5\times 10^9, 9.8\times 10^9, 1.7\times 10^{10},9.7\times 10^{10}$ yr.

\begin{figure}
\plotone{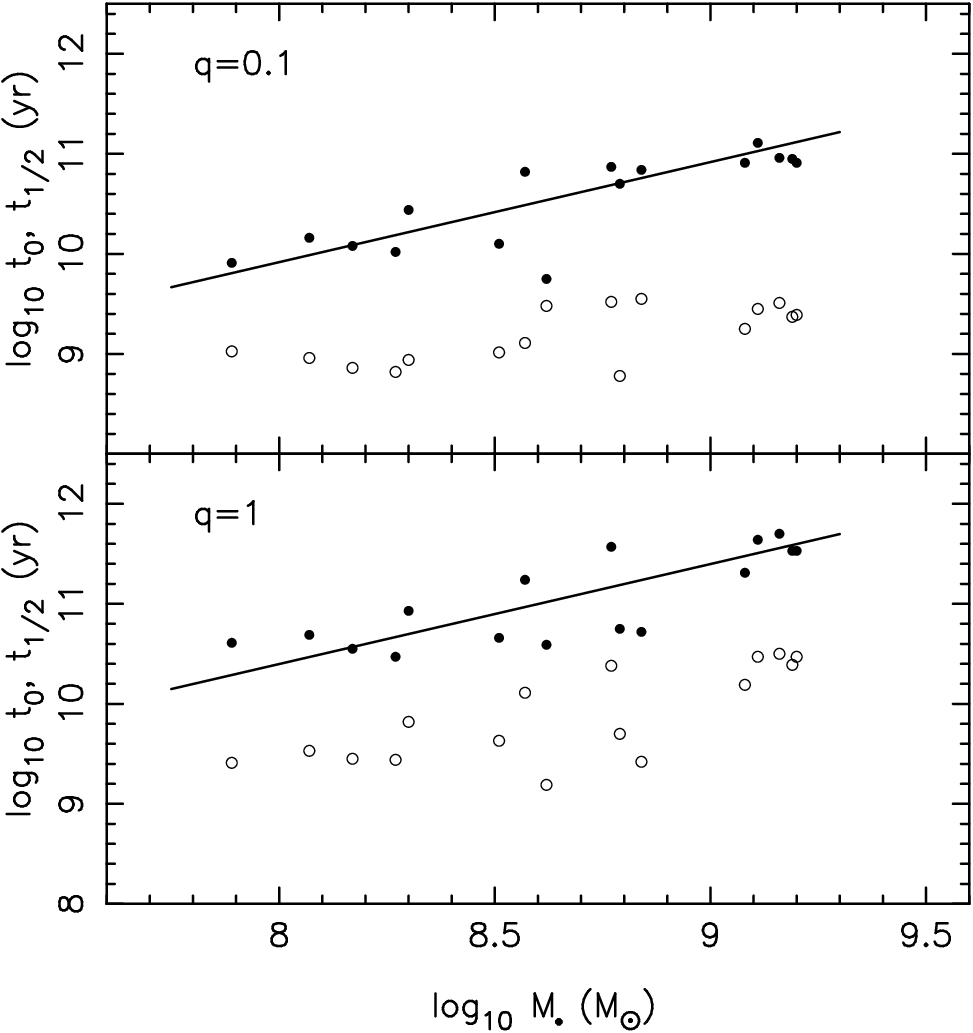}
\caption{Two characteristic times associated with loss-cone
refilling in the sample galaxies.
$t_0$ is the elapsed time before the first star is captured
and $t_{1/2}$ is the time for the loss-cone flux to reach
$1/2$ of its steady-state value.
Solid lines are the approximate fitting function,
equation (\ref{eq:fit}).
\label{fig:times}}
\end{figure}

We defined two characteristic times associated with loss-cone
refilling.
$t_0$ is the elapsed time before a single star
would be scattered into the loss cone. 
Obviously, the value of $t_0$ might depend rather sensitively
on the choice of initial conditions.
A more robust measure of the refilling time is $t_{1/2}$,
the elapsed time before the loss-cone flux reaches $1/2$ of
its steady-state value.
Figure~\ref{fig:times} shows values of $t_0$ and $t_{1/2}$ computed for
our sample of galaxies, using $K=1$ and $q=(0.1,1)$.
$t_0$ is of order $10^9$ yr or greater for all galaxies and for
both values of $q$, and $t_{1/2}$ is roughly an order of
magnitude greater.
The relation:
\beq
{t_{1/2}\over 10^{11}{\rm yr}} \approx C{\mu\over 10^8\msun} \approx
C {q\over (1+q)^2}{M_\bullet\over 10^8\msun}
\label{eq:fit}
\eeq
fits the data in Figure~\ref{fig:times} tolerably
well with $C= 1$.

\section{Discussion}

The flux of stars into the tidal disruption
loss cone of a BH that formed via binary coalescence
can be much lower than would be predicted under 
the steady-state assumption
\citep{su-99,mt-99,wang-04}.
Time scales for loss-cone refilling in the nuclei of bright
elliptical galaxies are of order $10^{10}$ yr
or longer and scale roughly linearly with BH mass for a fixed
(assumed) mass ratio of the binary that preceded the current,
single BH.
Even if the galaxy merger occurred as long as $\sim 5\times 10^9$
yr ago, the flux of stars into the BH's tidal disruption sphere could
still be less than $\sim 1\%$ of the steady-state value.
BH feeding rates in bright elliptical galaxies were already
expected to be lower  ($\lap 10^{-5}$ stars yr$^{-1}$; Paper I) 
than in less luminous
galaxies, due primarily to their low central densities.
Based on the arguments in this paper, stellar disruption rates in these
galaxies might be much lower still, making the nuclei of
bright ellipticals 
unlikely sites for observing a tidal flaring event.
Event rates in fainter galaxies, on the other hand,
should be closer to the steady-state values computed in 
Paper I since these galaxies 
are less likely to have experienced a merger since the epoch at which the
stellar cusp was formed, and since relaxtion times in their dense
nuclei are probably relatively short \citep{lauer-98}.

We have investigated an extreme, but physically plausible,
model in which the region around a newly-coalesced BH
was emptied of stars out to radii far greater than
the radius of the tidal disruption sphere.
But even in less extreme nuclear formation models, 
the long relaxation times in stellar nuclei imply a slow
approach to a steady-state distribution of stars near the
loss cone and hence a loss cone flux that can be very
different than the value computed via the
standard, steady-state theory.
Our results highlight the need for a fully 
time-dependent theory of loss cone evolution that can be applied to systems,
like galactic nuclei, that are much less than one relaxation time old.

\acknowledgments
This work was supported by grants 
AST-0071099, AST-0206031, AST-0420920 and AST-0437519 from the 
NSF, grant 
NNG04GJ48G from NASA,
and grant HST-AR-09519.01-A from
STScI.

\end{document}